\documentstyle[prl,aps,psfig]{revtex}
\def\Approx#1{{\mathstrut_{\displaystyle {\sim} }^{\displaystyle #1}}} 
\begin{document} 
\draft
\twocolumn[\hsize\textwidth\columnwidth\hsize\csname
@twocolumnfalse\endcsname

\title{Composite fermions close to the one-half filling of the lowest 
Landau level revisited}   

\author{Behnam Farid}

\address{Cavendish Laboratory, Department of Physics, 
University of Cambridge,\\
Madingley Road, Cambridge CB3 0HE, United Kingdom}

\date{\today}

\maketitle

\begin{abstract}
\leftskip 54.8pt
\rightskip 54.8pt
By strictly adhering to the microscopic theory of composite fermions 
(CFs) for the Landau-level filling fractions $\nu_{\rm e} = p/(2 p + 1)$, 
we reproduce, with remarkable accuracy, the surface-acoustic-wave 
(SAW)-based experimental results by Willett and co-workers concerning 
two-dimensional electron systems with $\nu_{\rm e}$ close to $1/2$. Our 
results imply that the {\sl electron} band mass $m_b$, as distinct from 
the CF mass $m_{\star}$, must undergo a substantial increase under the 
conditions corresponding to $\nu_{\rm e} \approx 1/2$. We further 
establish that a finite mean-free path $\ell_0$ is essential for the 
{\sl observed} linearity of the longitudinal conductivity $\sigma_{xx}(q)$
as deduced from the SAW velocity shifts.
\end{abstract}

\pacs{PACS numbers: 71.10.Pm, 73.40.Hm, 73.50.Jt} ]
\narrowtext
The fractional quantum Hall effect (FQHE) owes its existence to the 
electron-electron (e-e) interaction. The fermionic Chern-Simons field 
theory in $2+1$ dimensions unifies the FQHE with the integer QHE (IQHE) 
whose {\sl existence} does not depend on e-e interaction 
\cite{Dai,Lopez,Halperin}. This is effected through the binding, 
brought about by the mediation of the Chern-Simons action, of $2 n$, 
$n=1,2,\dots$, magnetic flux quanta to electrons, whereby the composite 
particles, that is CFs \cite{Jain}, are exposed to an effective magnetic 
flux density $\Delta B$ whose corresponding IQH state determines the 
FQH state for the electrons; the FQH state associated with $\nu_{\rm e} 
= p/(2 n p + 1)$ is the IQH state of CFs in which $p$ lowest Landau 
levels (LLs) are fully occupied. The sequence $\nu_{\rm e} = p/(2 n p 
+ 1)$ and its particle-hole conjugate $\nu_{\rm e}' {:=} 1 - p/(2 n p 
+ 1)$ approach $\nu_{\rm e} = 1/(2 n)$ for $p\to \infty$. In this work 
we deal with the case where $n=1$ and states whose $\nu_{\rm e}$ are 
close to $1/2$. With $\Delta B \equiv h n_e/(e p)$, where $-e < 0$ 
stands for the charge of an electron and $n_e$ the planar number density 
of electrons, these states thus correspond to small $\Delta B$. The state 
corresponding to $\nu_{\rm e}=1/2$ was proposed by Halperin, Lee and 
Read \cite{Halperin} to be a compressible state of degenerate fermions 
which, in the case of full polarisation of the electron spins, is 
characterised by a cylindrical Fermi surface of radius $k_F^{\sc cf} = 
\sqrt{4\pi n_e}$. This has been borne out by several experiments 
\cite{Willett1,Willett2,Kang,Goldman,Smet1}.

In this work we particularly concentrate on a series of experimental
results by Willett and co-workers \cite{Willett1,Willett2,Willett3},
concerning 2DESs with $\nu_{\rm e}$ at and close to $1/2$ and establish 
that these can be remarkably accurately reproduced within the framework 
of the microscopic Chern-Simons field theory. To achieve this, it turns 
out to be essential that the band-electron mass $m_b$ be by one order 
of magnitude larger than the customarily-assumed value: for GaAs 
heterostructures, in which the 2DESs under consideration were realised, 
$m_b$ is customarily taken to be $0.067\times m_e$, where $m_e$ is the 
electron mass in vacuum. In what follows, we use the notation $m_b^0 = 
0.067 \times m_e$ and denote the electron band mass, as required for 
reproducing the indicated experimental results, by $m_b$. Since quantum 
fluctuations, with respect to the mean-field approximation for CFs give 
rise to mass renormalisation \cite{Halperin}, we reserve $m_{\star}$ to 
denote the renormalised CF mass. Our numerical results imply $m_b 
\Approx> 16 \times m_b^0$ and $m_{\star} \approx 0.5 \times m_b \Approx> 
8 \times m_b^0 \approx 0.54 \times m_e$. This value is in good accord 
with the CF mass as deduced both from the values of the energy gaps 
$\Delta_{\nu_e}$ separating the CF LLs (determined from the activated 
temperature dependence of the longitudinal resistivities, $\varrho_{xx}$, 
centred around $\nu_e$'s close to $1/2$) \cite{Du1,Du2,Manoharan,Willett3} 
and the amplitude of the oscillations in $\varrho_{xx}$, for varying 
$\Delta B$, viewing these as the Shubnikov-de Haas oscillations 
corresponding to CFs \cite{Leadly,Du3,Coleridge,Manoharan,Willett3}. 
For completeness, according to Park and Jain \cite{Park}, the scale of 
the low-energy excitations of a 2DES may be determined by either the 
`activation mass' $m_a$ or the `polarisation mass' $m_p$, the latter being 
often the case for $\nu_{\rm e} > 1$. For $\nu_{\rm e} =1/2$ ($n_e=1.26
\times 10^{11}$~cm$^{-2}$) Kukushkin, {\sl et al.} \cite{Kukushkin} 
have reported a CF mass equal to $2.27 \times m_e$ which the authors 
suggest to be the $m_p$ of CFs. In view of the fact that the experiments 
in Ref.~\cite{Kukushkin} are based on optical excitations, involving 
$q\approx 0$, we suggest the possibility that the mass measured by 
these authors may be not $m_p$, but $m_b$, consistent with the 
requirement of a Kohn's theorem \cite{Kohn} (note the aspect 
$q \approx 0$).
 
In the experiments by Willett, {\sl et al.} 
\cite{Willett1,Willett2,Willett3}, the relative change $\Delta v_s/v_s$ 
in the velocity $v_s$ of SAWs as well as the damping $\kappa$ in their 
amplitudes, while propagating on the surface of samples at distance $d$ 
from the 2DES, were measured; here the change $\Delta v_s$ is relative to 
the $v_s$ corresponding to the case where the conductivity of the 2DES 
is infinitely large (see further on). Theoretically, $\Delta v_s/v_s$ is 
determined by the `on-the-mass-shell' value of the longitudinal 
conductivity $\sigma_{xx}(q,\omega)$ of the 2DES as follows
\begin{equation}
\label{e1}
\frac{\Delta v_s}{v_s} = \frac{\alpha^2/2}{1 + 
[\sigma_{xx}(q)/\sigma_m]^2},
\end{equation}
where $\sigma_{xx}(q) \equiv \sigma_{xx}(q,\omega=v_s q)$, $\sigma_m$ 
is a constant to be specified below, and \cite{Simon1} $\alpha^2/2 =$ 
$(e_{14}^2\tilde\epsilon_r/\{H \epsilon_0 \epsilon_r^2\})$ $\vert F(q d)
\vert^2$; here $e_{14} \approx 0.145$~Cm$^{-2}$ is the piezoelectric 
constant for Al$_x$Ga$_{1-x}$As, $\tilde\epsilon_r = \epsilon_r \big(1 + 
\{(\epsilon_r - 1)/(\epsilon_r + 1)\} \exp(-2 q d)\big)^{-1}$ ({\sl c.f.} 
Eq.~(A4) in Ref.~\cite{Simon1}), $H\approx 28.8\times 10^{10}$~Nm$^{-2}$,
$\epsilon_0 = 8.854\dots\times 10^{-12}$~Fm$^{-1}$ denotes the vacuum 
permittivity, $\epsilon_r$ is the relative dielectric constant of the 
{\sl bulk} of the host material which we take to be equal to $12.4$, 
and $F(x) {:=} A_1 \exp(-a x) \sin(b x + c) + A_2 \exp(-x)$ ({\sl c.f.} 
Eq.~(51) in Ref.~\cite{Simon1}), where $A_1 \approx 3.18$, $a \approx 
0.501$, $b \approx 0.472$, $c\approx 2.41$ and $A_2 \approx -3.10$; further,
\begin{equation}
\label{e2}
\sigma_m {:=} \left. \frac{\omega e^2}{q^2 v(q)}\right|_{\omega
= v_s q} \equiv 2 \epsilon_0 \tilde\epsilon_r v_s,
\end{equation}
where in the second expression on the right-hand side we have replaced 
the e-e interaction function $v(q)$ by the Coulomb function $v(q) = 
e^2/(2 \epsilon_0\tilde\epsilon_r q)$; we have further made use of the 
dispersion of acoustic phonons and employed $\omega= v_s q$, where $v_s$ 
stands for the sound velocity which in GaAs amounts to $3010$~ms$^{-1}$.

Our calculations are based on \cite{Halperin} (see also Eq.~(208) in 
Ref.~\cite{Simon2}) $\Delta v_s/v_s + i \kappa/q = (\alpha^2/2) 
\big( \gamma - v(q) K_{00}(q,\omega= v_s q)\big)$, where 
$-K_{00}(q,\omega)$ describes the change in $n_e$ to linear order in 
the {\sl external} potential \cite{Dai,Lopez,Halperin,Simon3}; the minus 
sign here has its origin in our convention with regard to the sign of $e$. 
In the literature, the constant $\gamma$ is invariably identified with 
unity, this on account of the fact that $\Delta v_s$ is the deviation of 
the measured $v_s$ with respect to the $v_s$ pertaining to the case where 
$\sigma_{xx}(q) \to \infty$. We maintain $\gamma$ in our considerations, 
following the fact that, experimentally, the reference $v_s$ does {\sl not} 
correspond to an infinitely large $\sigma_{xx}(q)$ (owing to impurities,
and $q\not\approx 0$).

In the present work we employ the `modified random phase approximation'
(MRPA) for $K_{00}(q,\omega)$ due to Simon and Halperin \cite{Simon3}
which takes account of the renormalisation of the mass of CFs and 
which approaches the RPA for $K_{00}(q,\omega)$ as $q\to 0$, the 
latter coinciding to leading order (proportional to $q^2$) with the 
{\sl exact} $K_{00}(q,\omega)$ for $q\to 0$ {\sl and} any $\omega$ 
at which both $K_{00}^{\sc rpa}(q,\omega)$ and $K_{00}(q,\omega)$ 
are bounded, which is the case for $\vert \omega\vert < \Delta\omega_c 
{:=} e \Delta B/m_b$ (see in particular Eq.~(5.6) in Ref.~\cite{Zhang}). 
The $K_{00}^{\sc mrpa}$ has thus the property that it conforms with 
the requirement of a Kohn's theorem \cite{Kohn} according to which 
$K_{00}(q\to 0,\omega)$ must be determined by $m_b$ rather than 
$m_{\star}$. For the explicit expression concerning $K_{00}^{\sc mrpa}
(q,\omega)$ corresponding to $\nu_{\rm e} = p/(2 n p + 1)$ in terms 
of elementary functions, we refer the reader to Ref.~\cite{Simon3} (we 
have presented and employed $K_{00}^{\sc mrpa}(q,\omega=0)$ in 
Ref.~\cite{Farid1}). We mention that for the purpose of calculating 
$K_{00}^{\sc mrpa}(q,\omega)$, which depends on $p$ and $n$, for a 
continuous range of $\Delta B$ around zero, we first determine 
$\nu_{\rm e}$ from $\nu_{\rm e} = h n_e/(e B)$, with $B$ the {\sl total} 
applied magnetic flux density, and subsequently obtain the required $p$ 
from $p = [\nu_{\rm e}/(1 - 2 \nu_{\rm e})]$ if $\nu_{\rm e} < 1/2$, 
and $p = [(1-\nu_e)/(1 - 2 (1-\nu_{\rm e}))]$ if $\nu_{\rm e} > 1/2$, 
where $[x]$ denotes the greatest integer less than or equal to $x$. The 
$\nu_{\rm e}$ corresponding to the thus-obtained $p$ remains constant 
for certain ranges of values of $B$, which causes artificial 
stepwise-constant behaviour in the functions of $\Delta B$ that depend 
on $K_{00}^{\sc mrpa}(q,\omega)$.

We model the effects of the impurity scattering through substituting 
$\omega +i/\tau_0$ for $\omega$ in $K_{00}^{\sc mrpa}(q,\omega)$. 
Here $\tau_0$ stands for the scattering time which is related to the 
mean-free path $\ell_0 = v_F \tau_0$, where $v_F$ stands for the Fermi 
velocity. In general, the substitution $\omega\rightharpoonup \omega + 
i/\tau_0$ amounts to solving the Boltzmann equation within the framework 
of the relaxation-time approximation, neglecting the so-called 
current-conservation correction which has been found to have {\sl no} 
significant consequences in contexts similar to that of our present 
considerations (see Sec. 4.1.4 in Ref.~\cite{Simon2}). With (from here 
onwards we suppress `{\sc mrpa}') $K'_{00}(q,\omega+i/\tau_0) \equiv 
{\rm Re}\big\{K_{00}(q,\omega +i/\tau_0)\big\}$ and $\gamma' {:=} 
{\rm Re}\{\gamma\}$, from the above-presented expression for $\Delta 
v_s/v_s + i \kappa/q$ we obtain
\begin{equation}
\label{e3}
\frac{\Delta v_s}{v_s} = \frac{\alpha^2}{2}
\big( \gamma' - v(q) K_{00}'(q,v_s q + i/\tau_0)\big).
\end{equation}
We eliminate $\gamma'$, whose value has {\sl no} influence on the {\sl 
form} of $\Delta v_s/v_s$, by requiring that $\Delta v_s/v_s$, as a 
function of $\Delta B$, coincide with the experimental $\Delta v_s/v_s$ 
for $\Delta B\approx 0$. 

In Fig.~1 we present $\Delta v_s/v_s$ as a function of $\Delta B$ 
for the cases where $n_e = 6.92\times 10^{10}$~cm$^{-2}$, $f {:=} 
\omega/(2\pi) = 8.5$~GHz (left panel), to be compared with Fig.~4 in 
Ref.~\cite{Willett2}, and $n_e = 1.6\times 10^{11}$~cm$^{-2}$, $f 
= 10.7$~GHz (right panel), to be compared with Fig.~1 in 
Ref.~\cite{Willett3}. The excellent agreement between the theoretical 
results corresponding to an enhanced $m_b$ (with respect to $m_b^0$) and 
experimental results, in particular when these are compared with those 
obtained within the same theoretical framework in which $m_{\star}$ 
retains the same enhanced value as compared with $m_b^0$ but $m_b
=m_b^0$ (curves (c)), strongly support the viewpoint that under the 
conditions where $\nu_{\rm e} \approx 1/2$, the bare band mass $m_b^0$ 
should be enhanced. This observation is compatible with the experimental 
finding with regard to the stronger than the theoretically-predicted 
divergence \cite{Halperin} of the CF mass for $\nu_{\rm e}\to 1/2$ 
\cite{Du2,Manoharan,Du3,Coleridge}. In this connection we should 
emphasise that the closer one approaches $\nu_{\rm e}=1/2$, the less 
sensitive $\Delta v_s/v_s$ becomes with respect to the further increase 
of $m_{\star}$ (or $m_b$ for that matter); our choices $m_b=16 \times 
m_b^0$ and $m_{\star} = 0.5\times m_b$ are based on the consideration 
that the experimental features corresponding to $\Delta B$ in the range 
$\sim 0.1 - 1$~T be reproduced. The results in Fig.~1 obtained through 
employing the semi-classical $\sigma_{\mu\nu}$, due to Cohen, Harrison 
and Harrison (CHH) \cite{Cohen} (see also Appendix B in 
Ref.~\cite{Halperin} as well as Eqs.~(2) and (3) in Ref.~\cite{Willett2}), 
bring out the inadequacy of the semi-classical approach; curves marked 
by (d) unequivocally demonstrate the shortcoming of strictly adhering 
to the viewpoint that CFs behaved like non-interacting electrons exposed 
to a reduced magnetic field --- curves marked by (b), which are similarly 
based on the CHH $\sigma_{\mu\nu}$, owe their resemblance to the 
experimental results to the fact that in their calculation {\sl explicit} 
account has been taken of the conditions which are specific to the regime 
corresponding to $\nu_{\rm e}\approx 1/2$. With reference to our earlier 
work \cite{Farid1}, it is appropriate to compare curve (a) in the right 
panel of Fig.~1 with the curves in Fig.~1 of the work by Mirlin and 
W\"olfle \cite{Mirlin} and compare both with the experimental trace in 
Fig.~1 of Ref.~\cite{Willett3}. One observes that our present result, in 
contrast with those in Ref.~\cite{Mirlin}, precisely reproduces almost 
{\sl all} features of the experimental trace, such as the values of 
$\Delta v_s/v_s$ at $\Delta B \approx 0, 0.38, 0.54, 1.0$~T.

From Eqs.~(\ref{e1}) and (\ref{e3}) we obtain
\begin{equation}
\label{e4}
\frac{\sigma_{xx}(q)}{\sigma_m} = \Big(\big(\gamma'
- v(q) K'_{00}(q,v_s q+i/\tau_0)\big)^{-1} -1\Big)^{1/2},
\end{equation}
according to which $\sigma_{xx}(q)$ interestingly does {\sl not} 
explicitly depend upon $\alpha^2/2$. Unless we set $\gamma'=1$, we 
eliminate $\gamma'$ in Eq.~(\ref{e4}) by requiring that for given 
values of $\sigma_m$ and $q$, $\sigma_{xx}(q)$ according to 
Eq.~(\ref{e4}) yield the corresponding experimental SAW velocity shift.
\begin{figure}[t!]
\protect
\label{fi1}
\centerline{
\psfig{figure=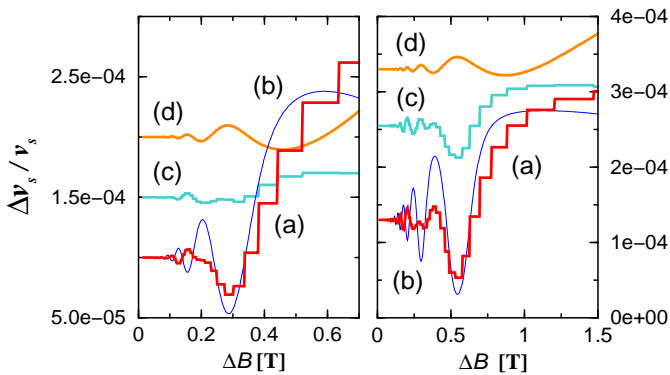,width=3.5in} }
\vskip 10pt
\caption{\rm (colour)
The relative shift of the SAW velocity. (a) and (c) are obtained from 
Eq.~(\protect\ref{e3}), while (b) and (d) are based on 
Eq.~(\protect\ref{e1}) in which the semi-classical result for 
$\sigma_{\mu\nu}$, due to CHH \protect\cite{Cohen}, has been employed: 
for (d), $\sigma_{xx}(q)$ has been directly used, whereas for (b) 
$\sigma_{xx}(q)$ has been obtained from $\sigma_{yy}(q)$ according to 
$\sigma_{xx}(q) \approx (\nu_e e^2/h)^2\vert_{\nu_e=1/2}/\sigma_{yy}(q)$ 
--- this expression takes explicit account of the fact that for $\nu_e 
\approx 1/2$, $\varrho_{xx} \varrho_{yy} \ll \varrho_{xy}^2$ and 
$\varrho_{yy} \approx 1/\sigma_{yy}$; in both cases we have scaled 
$\sigma_{xx}/\sigma_m$ by a constant such that for $\nu_{\rm e}=1/2$ 
the corresponding $\Delta v_s/v_s$ coincide with the experimental value 
(for clarity we have offset (d) by $10^{-4}$ in the left panel and by 
$2\times 10^{-4}$ in the right). The equality of (a) and (c) (we 
have offset (c) by $5\times 10^{-5}$ in the left panel and by 
$1.75\times 10^{-4}$ in the right) with the associated experimental 
results at $\nu=1/2$ has been enforced through appropriate choices 
for $\gamma'$ ($\approx 1.1$).
{\it Left panel}: 
$n_e = 6.92 \times 10^{10}$~cm$^{-2}$, $\ell_0 = 0.4$~$\mu$m, 
$d=10$~nm, $f=8.5$~GHz.
{\it Right panel}: 
$n_e = 1.6 \times 10^{11}$~cm$^{-2}$, $\ell_0 = 0.6$~$\mu$m
(corresponding to $\tau_0 = 19.6$~ps), 
$d=100$~nm, $f=10.7$~GHz.
{\it Both panels}: 
(a), $m_b = 16 \times m_b^0$, $m_{\star} = 0.5 \times m_b$;
(c), $m_b = m_b^0$, $m_{\star} = 8 \times m_b$.
For the origin of the step-like behaviour in curves (a) and (c) see 
the main text.}
\end{figure}
In Fig.~2 we present our theoretical $\sigma_{xx}(q)$ in comparison 
with its SAW-derived experimental $\sigma_{xx}(q)$ by Willett, {\sl 
et al.} \cite{Willett1} (see the middle panel of Fig.~2 herein). The 
details in Fig.~2 again support our above finding with regard to $m_b$ 
and $m_{\star}$, that a mere enhancement of $m_{\star}$ with respect 
to $m_b^0$ is not sufficient (see inset [B]). We also observe that a 
finite $\ell_0$ is most crucial to the experimentally-observed linear 
behaviour of the SAW-deduced $\sigma_{xx}(q)$ for $q$ in the range 
$\sim (0.015/a_0, 0.075/a_0$), with $a_0$ the Bohr radius (see curve 
(c)); the original observation with regard to $\sigma_{xx}(q) \propto q$ 
for $q \gg 2/\ell_0$ \cite{Halperin} thus turns out to be relevant for 
values of $q$ far outside the experimental range. We note that $\ell_0 
\approx 0.5$~$\mu$m coincides with that reported in the pertinent 
experimental articles. A further aspect that our present results in 
Fig.~2 clarify is that, in contrast to earlier observations (see the 
paragraph following Eq.~(7.6) in Ref.~\cite{Halperin} and that following 
Eq.~(211) in Ref.~\cite{Simon2}), the available experimental results by 
{\sl no} means are in conflict with the predictions of Eq.~(\ref{e2}): our 
theoretical results for $\sigma_{xx}(q)$ in Fig.~2 have been obtained 
through multiplying $\sigma_{xx}(q)/\sigma_m$ by {\sl the same} 
\cite{Willett4} $\sigma_m$ that has been employed to determine 
$\sigma_{xx}(q)$ from the SAW-deduced $\sigma_{xx}(q)/\sigma_m$. Thus, 
rather than Eq.~(\ref{e2}) being inadequate, the empirical method of 
determining $\sigma_m$ (which employs the {\sl dc} conductivity 
\cite{Willett3,Willett4}) should be considered as inappropriate.

We now briefly focus on the physical significance of the expression 
for $\Delta v_s/v_s$ in Eq.~(\ref{e1}). To this end, let 
$\delta v_{\rm ext}({\bf r},t)$ denote an applied time-dependent 
external potential, representing that corresponding to the SAWs. The 
change in the Hamiltonian of the system, following the application of 
$\delta v_{\rm ext}$, has the form $\delta \widehat{H}(t) = \int 
{\rm d}^2r\, \delta v_{\rm ext}({\bf r},t) \hat{\psi}^{\dag}({\bf r}t) 
\hat{\psi}({\bf r}t)$, where $\hat{\psi}^{\dag}({\bf r}t)$, 
$\hat{\psi}({\bf r}t)$ denote creation and annihilation field operators. 
Denoting the change in the energy of the system corresponding to 
$\delta \widehat{H}(t)$  by $\delta E(t)$, we have \cite{Farid2} 
$\delta E(t) =$ $\int {\rm d}^2r$ $\delta v_{\rm ext}({\bf r},t) 
\bar{n}_e({\bf r},t)$, where $\bar{n}_e({\bf r},t) {:=} \int_0^1 
{\rm d}\lambda\, n_e^{(\lambda)}({\bf r},t)$. Here $n_e^{(\lambda)}
({\bf r},t)$ stands for the instantaneous number density of the system 
corresponding to $\delta v_{\rm ext}^{(\lambda)}({\bf r},t) {:=} \lambda 
\delta v_{\rm ext}({\bf r},t)$. By assuming $\delta v_{\rm ext}({\bf r},t) 
\equiv \delta v_{\rm ext}({\bf r}) \exp(-i\omega_0 t)$, one obtains for 
$\delta {\cal E}(\omega) {:=} \int {\rm d}t$ $\delta E(t) \exp(i\omega t)$, 
$\delta {\cal E}(\omega) = -\int {\rm d}^2r {\rm d}^2r'\, \delta 
v_{\rm ext}({\bf r}) \overline{K}_{00}({\bf r},{\bf r}'; \omega-\omega_0) 
\delta v_{\rm ext}({\bf r}')$, where $\overline{K}_{00}({\bf r},{\bf r}';
\omega) {:=} \int_0^1 {\rm d}\lambda\, (1-\lambda)\, K^{(\lambda)}
({\bf r},{\bf r}';\omega)$ \cite{Farid2}. Under the {\sl assumption} 
that $\delta v_{\rm ext}({\bf r})$ be weak, the dependence upon $\lambda$ 
of $K^{(\lambda)}_{00}$ can be neglected so that $\overline{K}_{00} 
\approx 2^{-1} K^{(0)}_{00}$ where $-K^{(0)}_{00}$ stands for the 
density-density response function of the uniform, unperturbed, system. 
To second order in $\delta v_{\rm ext}(q)$, one thus obtains $\delta 
{\cal E}(\omega) = -2^{-1} K_{00}(q,\omega-\omega_0) \vert \delta 
v_{\rm ext}(q)\vert^2$. Let now $\overline{\delta E}_t {:=} \int_0^{t} 
{\rm d}\tau\, \delta E(\tau)$, from which one readily obtains 
$\overline{\delta E}_t \approx -2^{-1} \vert \delta v_{\rm ext}(q)\vert^2 
\int {\rm d}\omega\, K_{00}(q,\omega-\omega_0) \{1-\exp(-i\omega t)\}/(2\pi 
i\omega)$. For large $t$, the integrand of the $\omega$ integral becomes 
highly oscillatory so that to the leading order in $1/t$, $\int 
{\rm d}\omega\, K_{00}(q,\omega-\omega_0) \{1-\exp(-i\omega t)\}/(2\pi 
i\omega) \sim K_{00}(q,-\omega_0) \int {\rm d}\omega\, \{1-\exp(-i\omega 
t)\}/(2\pi i\omega) = K_{00}(q,-\omega_0)$. Since $K_{00}(q,\omega)$ 
is an even function of $\omega$, we eventually obtain $\overline{\delta 
E}_t \sim -2^{-1} K_{00}(q,\omega_0) \vert\delta v_{\rm ext}(q)\vert^2$, 
for $t \gg 1/\omega_0$. This expression, which coincides with Eq.~(14) 
in Ref.~\cite{Simon1}, is the fundamental link between $K_{00}(q,\omega)$ 
and $\Delta v_s/v_s + i\kappa/q$ presented above. These details make 
explicit first, that only small-amplitude perturbations are correctly 
accounted for by Eq.~(\ref{e1}) (and similarly, Eq.~(\ref{e3})), and 
second, that the observation of ``geometric resonance'' and the 
``cyclotron frequency deduced from dc transport'' though, as suggested 
by Willett, {\sl et al.} \cite{Willett3}, inconsistent with ``a 
non-interacting, semi-classical quasiparticle model [for CFs]'', are in 
fact {\sl not} inconsistent with the physical picture that the above 
derivation brings out: that the mechanism underlying the SAW-based 
experiments does not involve any resonance phenomenon in the usual sense 
and that the SAW experiments, which involve a long-time {\sl integration} 
of the fluctuations in the total energy of 2DESs, unveil $K_{00}(q,\omega)$ 
at $\omega =\omega_0 \equiv v_s q$ and $q=\omega_0/v_s$, {\sl independent} 
of the magnitude of the CF cyclotron frequency $\Delta \omega_{c\star}$ 
and consequently of that of the CF mass $m_{\star}$.

\begin{figure}[t!]
\protect
\label{fi2}
\centerline{
\psfig{figure=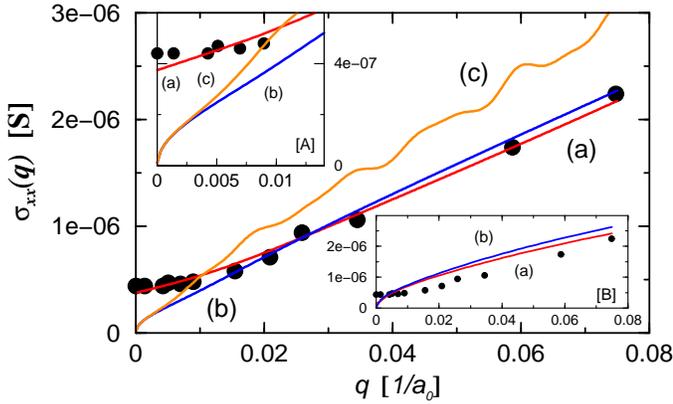,width=3.5in} }
\vskip 10pt
\caption{\rm (colour)
The longitudinal conductivity as deduced from the SAW $\Delta v_s/v_s$ 
at $\nu=1/2$ (here $a_0 = 9.79$~nm) corresponding to $n_e = 6.27 \times 
10^{10}$~cm$^{-2}$ ($k_F^{\sc cf} = 0.87/a_0$). Solid dots are 
experimental results by Willett, {\sl et al.} \protect\cite{Willett1}, and 
(a), (b) and (c) have been calculated according to Eq.~(\protect\ref{e4}); 
for all three cases we have assumed $m_b = 16\times m_b^0$, $m_{\star} 
= 0.6 \times m_b$, $d=10$~nm (for this value of $d$, $\sigma_m = 3.57
\times 10^{-7}$ ${=:} \sigma_m^{\rm th}$ according to 
Eq.~(\protect\ref{e2})); in order to compare our results with those in 
Ref.~\protect\cite{Willett1}, we have multiplied the theoretical 
$\sigma_{xx}(q)/\sigma_m$ by the value for $\sigma_m$ employed in this 
reference, namely \protect\cite{Willett4} $\sigma_m = 1.67 \times 
10^{-6}$~S. (a) has been determined with $\ell_0 = 0.6$~$\mu$m, (b) with 
$\ell_0 = 0.9$~$\mu$m and (c) with $\ell_0 = 2.4$~$\mu$m. (b) and (c) have 
been calculated with $\gamma'=1$, while (a) has been obtained by following 
the procedure outlined in the text: taking the experimental result 
$\sigma_{xx}(q=0.004/a_0) = 0.44 \times 10^{-6}$~S, we have obtained and 
used $\gamma'=0.952$. Inset [A] is a focus on the small-$q$ region of the 
main diagram (the $q^{1/2}$-behaviour of (b) and (c) for $q\to 0$ is 
associated with $v(q) K_{00}(q,\omega) \propto q$ in this region), while 
inset [B] shows the counterparts of curves (a) and (b) for $m_b = m_b^0$, 
$m_{\star} = 8 \times m_b$ (for (a), $\gamma' =1.008$). }  
\end{figure}
In conclusion, by strictly adhering to the microscopic theory of CFs, 
we have established that the SAW-based experimental results by Willett 
and co-workers close to $\nu_{\rm e} =1/2$ can be remarkably accurately 
reproduced provided the electron band mass $m_b$ be substantially
enhanced with respect to $m_b^0$; in this picture, the (observed) large 
value of the CF mass $m_{\star}$ follows from a subsequent reduction 
of $m_b$ (owing to quantum fluctuations) rather than a direct enhancement 
of $m_b^0$. We have further established that a finite mean-free path 
$\ell_0$ is essential to the experimentally-observed linearity in the 
SAW-deduced $\sigma_{xx}(q)$ in the range $1.56 \Approx< q \Approx< 
7.68$ $\mu$m$^{-1}$, and that there exists {\sl no} discrepancy 
between the theoretical and experimental values for $\sigma_m$.

I thank Professors D.~E. Khmel'nitski\v{i} and P.~B. Littlewood for 
a discussion and Professor B.~I. Halperin and Drs S.~H. Simon and 
R.~L. Willett for kindly clarifying some aspects concerning the 
empirical $\sigma_m$. With appreciation I acknowledge hospitality of 
Cavendish Laboratory.

%

\end{document}